\documentstyle[12pt]{article}
\advance\textheight by 60pt
\advance\voffset by -40pt
\advance\textwidth by 50pt
\advance\oddsidemargin by -25pt
\advance\evensidemargin by -25pt

\def\single_space{\baselineskip 12pt plus 1pt minus 1pt}
\def\one_and_a_half_space{\baselineskip 19pt plus 1pt minus 1pt}
\def\double_spacesp{\baselineskip 25pt plus 2pt minus 2pt}
\begin{document}
\baselineskip 25pt plus 2pt minus 2pt
\thispagestyle{empty}
\begin{center}
{\Large $J/\psi$ at small-$x$\\}
\vspace*{0.2in}
M. A. Doncheski$^a$, M. B. Gay Ducati$^{a,b}$, F. Halzen$^a$ \\
\vspace*{0.2in}
$^a$Department of Physics, University of Wisconsin, Madison, WI 53706 \\
\vspace*{0.1in}
$^b$Instituto de F\'\i sica, Universidade Federal do Rio Grande do Sul \\
C. P. 15051, 91500, Porto Alegre, RS, Brazil \\
\vspace*{0.2in}
Abstract \\
\end{center}

It has been suggested that the suppression of $J/\psi$ production in
heavy nuclei is a signature of the formation of quark-gluon plasma.  We
here show that this phenomenon can be understood in terms of conventional
physics, {\it i.e.} {\it i)} perturbative QCD, {\it ii)} the parton
recombination implementation of shadowing in the initial state, and
{\it iii)} final state interactions with the hadronic debris of the
nuclear target.  Unlike previous calculations we include both the direct
$J/\psi$ production and its production via radiative $\chi$ decays
($\chi_J \to J/\psi + \gamma$).  We are able to reproduce the
experimental data including their small-$x$ behavior.  We emphasize the
importance of studying the $x_2$-dependence of the ratio
$\sigma(b  A)/\sigma(b N)$, where $b$ designates the beam and $x_2$ is
the momentum fraction of the parton from the nuclear target.

\newpage

\section{Introduction}

It has been suggested that the suppression of $J/\psi$ production on
heavy nuclei is a signal for the formation of the quark-gluon plasma;
however we would like to consider a conventional physics solution to the
problem.  $J/\psi$ production will be described by conventional
perturbative QCD after including the modification of the small-$x$
behavior of the distribution functions for gluons and sea quarks in the
nuclear medium~\cite{mueller}. This allows for the continued use of the
hard scattering amplitudes and the factorization theorem.  This
modification, plus a correction for the `classical EMC' effect at
intermediate $x$, provides a multiplicative modification of the parton
distribution functions, for partons in a nuclear target. Furthermore,
because of the hadronic nature of the final state, effects of the
interaction of the final state with the hadronic debris of the nuclear
target must be included. We will show how such a picture can reproduce in
detail a wide range of data on lepto- and hadroproduction of $J/\psi$.

$J/\psi$, $\psi^{'}$ and $\Upsilon$ suppression on nuclear
targets~\cite{alde1,amaudruz1} provides valuable data for the study of
the $A$-dependence at small-$x$ region and/or large $x_{F}$.  Here $x$ is
the Bjorken-$x$ variable, the momentum fraction of a parton within a
hadron, and $x_F$ is the Feynman-$x$ variable, $x_F = x_1 - x_2$ where
$x_1$ and $x_2$ are Bjorken-$x$ variables for the partons from the beam
and target, respectively.  The experimental energy ranges from 40 to
800~GeV, which spans part of the intermediate $x$ range through the
small-$x$ region for $J/\psi$ production.

$J/\psi$ production in Deep Inelastic Scattering (DIS) has been measured
by the NMC~\cite{amaudruz1} collaboration for incident muon energies of
200 and 280~GeV.  The results are presented as a ratio of cross sections
for two different targets: Sn and C.  These data are relevant to the
study of the $A$-dependence of the cross-section at small $x$.  In fact,
$A$-dependence in $J/\psi$ production is already known for hadronic
processes~\cite{katsanevas} and the cross section ratio
\begin{equation}
          R=\sigma^{hA_{1}}/\sigma^{hA_{2}}
\end{equation}
has been studied considering initial and final state nuclear
effects~\cite{epele,close}.  This ratio is for $J/\psi$ production less
than unity in the small x region. This result cannot be explained by QCD
parton model, which gives the ratio equal to 1.

Nuclear effects have been observed in high energy processes at different
momentum transfer, $Q^{2}$.  Those effects can be traced in DIS with
neutrinos~\cite{allport} or charged lepton
beams~\cite{amaudruz1,arneodo,pamaudruz,arneodo2}, in Drell-Yan
processes~\cite{alde2}, and in hadroproduction of heavy
quarks~\cite{katsanevas,alde1,duffy,badier}.  Only the intensity of the
suppression differs, and this is a key aspect.  The general behavior is
$A_{eff}/A < 1$, where $A_{eff}$ is defined as $\sigma(bA)$/$\sigma(bN)$,
with $b$ designating the beam.  Common features include a more pronounced
effect for heavier target nuclei, a rapidly diminishing effect with
increasing $x$ and very little dependence on $Q^{2}$.  The first result
on nuclear dependence were obtained by EMC at intermediate
$x$~\cite{aubert}.  Since then the experimental results have been
extended to smaller $x$ values, and they exhibit the shadowing
phenomena.  The available data for $J/\psi$ production are presented in
Table 1.

\section{The model}

In DIS as well as in hadroproduction, the gluon fusion process is
dominant~\cite{ellis,berger} and the small-$x$ behavior of the target
gluon is crucial.  In order to understand the heavy meson suppression
with nuclear targets, it is critical to understand small-$x$ behavior of
the gluon structure function.  We here consider a recombination model to
take into account shadowing at the parton level in the initial state of
the process~\cite{mueller}.  This approach introduces a modification of
the parton evolution equations in order to take into account the
superposition probability when the partons have a large longitudinal size
(or large $1/x$).  This model incorporates the recombination through
ladder diagrams as a perturbative mechanism enabling a factorized
calculation for the cross section ratios~\cite{mueller}.  This approach
successfully explains both EMC and Drell-Yan small-$x$ data~\cite{ayala}.

The recombination effect is enhanced in the nuclear medium, where the
longitudinal size of the parton, $\Delta z$, can exceed the size of the
nucleon at small-$x$ region.  The quantity measured experimentally is a
ratio of structure functions,
\begin{equation}
R(x, Q^2, A) = \frac{F_2^A(x,Q^2)}{AF_2(x,Q^2)}
\end{equation}
and this quantity is found to be approximately $Q^2$ independent.  The
$x$- and $A$-dependence of this ratio has been parameterized by Berger
and Qiu~\cite{bq}, and these authors find that it factors in the DIS case:
\begin{equation}
R_{EMC} (x, Q^2, A) \approx R_g(x, A) R_a(x, A).
\end{equation}
The $R_g(x,A)$ factor is associated with the partonic shadowing, and it
has the functional form:
\begin{equation}
R_g (x,A) = \left\{ \begin{array}{ll}
                    1 & \mbox{$x_c < x < 1$} \\
                    1 - K_g (A^{1/3} - 1) \left[
                    \frac{\mbox{$\Delta z - \Delta z_c$}}
                    {\mbox{$\Delta z_A - \Delta z_c$}}
                    \right] & \mbox{$x_A < x < x_c$} \\
                    1 - K_g (A^{1/3} - 1) & \mbox{$0 < x < x_A$}
                    \end{array}
\right.
\end{equation}
where $K_g$ parameterizes the amount of gluon shadowing,
$\Delta z = 1/(x p)$ is the wavelength of the gluon,
$\Delta z_c =  1/(x_c p)$ is the longitudinal distance at which
neighboring nucleons begin to interact and $\Delta z_A = 1/(x_A p)$ is
the longitudinal size of the nucleus.  Thus the variables $x_A$ and $x_c$
are related to the Bjorken-$x$ corresponding to a probe of the nucleus
and nucleon, respectively.  We assume that $x_A = x_c/A^{1/3}$, following
eq.(9) of Ref.~\cite{bq}.  The remaining factor, $R_a(x, A)$,
parameterizes the classical EMC effect, and it has the approximate form:
\begin{equation}
R_a(x,A) = \frac{x}{x_1} + K_a (1 - \frac{x}{x_1}) \;,
\end{equation}
with this parameterization valid for $0 \leq x \leq 0.6$.

The above factors incorporate the `classical EMC' effect and initial
state effects. Final state effects are incorporated via a factor,
$R_{ss}(x_F,A) = A^{(\alpha (x_F) - 1)}$, where
$\alpha (x_{F})=0.97-0.27x_{F}^{2}$.  This expression is suggested by
data and also agrees with open charm results.  This should be adequate.
It is premature to deal with specific final state effects without having
a better understanding of the gluon distribution function behavior at
small-$x$.  Also, our parameterization of final state effects depends
only on a final state variable ($x_F$).

Other descriptions of final state interactions have been suggested. The
nuclear approach \`a la Glauber includes rescatterings of the $Q \bar{Q}$
or heavy meson in the nuclear medium~\cite{capella}.  Other authors also
consider specific final state effects (and attempts to make quantitative
predictions of them)~\cite{vogt} or Pomeron exchange
models~\cite{castorina}, or raise the fundamental question of the
validity of factorization in this kinematical region~\cite{collins,hoyer}.

\section{$J/\psi$ in DIS}

Deep inelastic photoproduction of $J/\psi$ is understood to take place by
the photon-gluon fusion mechanism~\cite{berger}
$\gamma + g_{1}\to J/\psi + g_{2}$.  The extra gluon in the final state
is required for the $c \bar{c}$ to be produced as a color singlet with
the correct quantum numbers ($J^P = 1^-$) of the $J/\psi$; color and spin
projection techniques were used to extract the relevant part of the
$\gamma + g \to c \bar{c} + g$ amplitude.  It has been shown that in the
inelastic region ($z=E_{J/\psi}/E_{\gamma}<0.9$) both gluons are hard and
it is not very relevant to take care of higher order multiple gluon
diagrams.


We combine the color singlet model for leptoproduction (photoproduction)
of the $J/\psi$  with the parton recombination model to account for the
$A$-dependence of the gluon function. We use the Weizs\"acker-Williams
approximation for the photon and the Morfin-Tung leading order set of
parton distributions~\cite{morfin} for the gluon.  An advantage of DIS
compared to hadroproduction in that there are no interactions with
hadronic components of the beam~\cite{vogt}.

The color singlet model for $J/\psi$ production reproduces the rapidity
distribution~\cite{allasia}. The color singlet model combined with the
Weizs\"acker-Williams approximation has been shown to agree with the
electroproduction of $J/\psi$ with a 15~GeV electron beam at
SLAC~\cite{ugo} and muoproduction with a 280~GeV muon beam at
CERN~\cite{emc,amaudruz1}, as long as experimental cuts ensure that the
$J/\psi$ production is inelastic.  We therefore feel confident in
applying the method to our study of shadowing in $J/\psi$ muoproduction
on nuclear targets. Also, the model has the interesting features of
providing a direct measure of the gluon distribution function, in the
case $\gamma N \to J/\psi+X$, or the direct determination of the gluon
$A$-dependence by means of Eq.~(1).

We assume that there is no EMC effect in the carbon target, since it is
light and for tin, the parameters $K_a$ and $x_1$ are 1.20 and 0.25
respectively.  We also choose a fixed value of $x_c$ for carbon, given by
Eq.~(9) in Ref.~\cite{bq},
which for $A = 12$ gives $x_c = 0.10$.  For tin, we allow $x_c$ and $K_g$
to be free parameters, and try to fit the existing data.  We choose the
ranges $0 \leq x_c \leq 0.25$ and $0 \leq K_g \leq 0.50$.  We then
calculate the ratio of the cross sections at the $x$ values of the NMC
data points, and perform a $\chi^2$ analysis. We now describe our results
for $A$-dependence on DIS and hadroproduction of $J/\psi$ ($\psi'$).

The DIS data is still not abundant but it is extremely useful for
comparison with the hadronic case.  Our results are presented in Fig.~1
and it is clear that more statistics are needed to enable a more critical
analysis of this model.  Due to the large error bars it is difficult to
constrain the parameters.  Also an extension to lower $x$ is important to
provide a better definition of the $x$ behavior of the ratio, Eqn. (2).

\section{$J/\psi$ in $h p (A)$}

The hadroproduction of $J/\psi$ can proceed through a number of parton
level processes.  The leading order source of $J/\psi$ in hadron-hadron
collisions is due to gluon-gluon fusion
($g + g \to J/\psi + g$)~\cite{Baier,Gastmans}, although it is not the
most improtant contribution.  The combination of a lower order (in
$\alpha_s$) and large branching fraction makes $\chi_J$ production,
followed by radiative decay ($\chi_J \to J/\psi + \gamma$), the dominant
source of $J/\psi$ (hereafter referred to as radiative production).  The
leading contribution to $\chi_J$ production is the low $p_T$ process
$g + g \to \chi_{0,2}$.  Additional contributions of the same order in
$\alpha_s$ as direct production come from $g + g \to \chi_{0,1,2} + g$,
$q + g \to \chi_{0,2} + q$ and $q \bar{q} \to \chi_{0,2} + g$.  All the
required parton level cross sections can be found, {\it e.g.}, in
Refs.~\cite{Baier,Gastmans}.  Furthermore, we use the Morfin-Tung leading
order parton distribution functions~\cite{morfin} for partons from the
target or from the proton beam, and Owens pion set 1~\cite{owens} for
partons from the pion beam.  We are attempting to perform a more careful
calculation of $J/\psi$ production on heavy nuclear targets than previous
analyses, and so the cross sections we use are those in which the correct
color singlet structure, the correct angular momentum quantum numbers and
small relative momenta are projected out.\footnote{Because of the
existence of $g + g \to \chi_{0,2}$, the higher order (non-zero $p_T$)
processes involving $\chi_{0,2}$ diverge at low $p_T$.  This is merely an
artifact of an incomplete calculation.  If one performs a full
calculation of $\chi_{0,2}$ ($g + g \to \chi_{0,2}$ including all 1 loop
graphs) production, and cancels the low $p_T$ divergences against virtual
infrared divergences and then absorbs the remaining collinear divergences
into the running of the parton distribution functions, the cross section
is indeed finite.  We adopt a less rigorous approach to this problem.  It
is known that at low $p_T$, $d \sigma / d p_T^2 \propto e^{-6 M_T}$ where
$M_T$ is the transverse mass of the charmonium state.  Also, the
divergences in $|A(gg \to \chi_{0,2} g)|^2$ and
$|A(qg \to \chi_{0,2} q)|^2$ are $1/t$, which gives $1/p_T^2$ at low
$p_T$.  Therefore we regularize these squared amplitudes with a factor
\begin{equation}
\left( \frac{p_T}{p_{T0}} \right)^3 e^{-6(M_T - M_{T0})}
\end{equation}
where $p_{T0}$ is a free parameter.  At low $p_T$, $|A|^2$ should be (a
slowly varying function of $p_T$)/$p_T^2$, and so our regularization
should reproduce the observed $p_T^2$ distribution.  To find a value for
$p_{T0}$, we first calculate the total $\chi_{0,2}$ cross section from
the $g + g \to \chi_{0,2}$ subprocesses and vary $p_{T0}$ until the
integration of the regularized $|A|^2$ yields the correct value.}  We
then include the factors $R_{EMC}$ and $R_{ss}$ in the calculation of the
various cross sections for production on the heavy nuclear target.  The
$J/\psi$ cross section and $x_2$ distribution can be constructed from the
direct and radiative $\chi_J$ components, with the inclusion of branching
ratios where appropriate.

As in the DIS case, we allow $x_c$ and $K_g$ to be free parameters, and
try to fit the existing data.  We choose the ranges
$0 \leq x_c \leq 0.25$ and $0 \leq K_g \leq 0.50$.  We then compare the
experimental data points on $R(x_2)$ (where $x_2$ is the parton momentum
fraction of the gluon from the target nucleus) to our calculations, and
perform a $\chi^2$ analysis.

We consider data from $p$ and $\pi$ beams with energies from 200 to
800~GeV, focusing on higher mass nuclear targets where the effect of a
heavy nucleus is more pronounced.  In Fig.~2 the results are presented
for $J/\psi$ (Fig.~2a) and $\psi'$ (Fig.~2b).  The best agreement
requires different parameters for each case.  This should not be
surprising since the $\psi^{'}$ is a spatially bigger resonance than
$J/\psi$.  A more refined version of this calculation should include this
fact in the final state effects.  We obtain very good agreement for
800~GeV for the $J/\psi$ case.  It is interesting to note that for the
$\psi'$ case, the value of $x_c$ required for the best fit is rather
large (the largest we allow in our analysis).  A reasonable fit to the
data can be obtained, in this case, in a range of parameter space as
demonstrated by the dashed line in Fig.~2b.

Also for 200~GeV and $p$ beam the agreement is good and the results are
shown in Fig.~3.  However for $\pi$ we reproduce quite well the general
behavior at 280 and 200~GeV, as shown in Figs.~4 and 5, respectively, but
for smaller $x$ our result is below the data.  It is likely that the pion
distribution function used is not suitable for this kinematical region.

As an overall result, considering that higher statistics are still needed
to clarify this complicated problem, we believe that a conventional model
as the one presented here is able to accommodate the data.  However, it
should be emphasized that a better understanding of the small-$x$
behavior of the gluon distribution function is needed, and in this HERA
and photoproduction experiments can play an important role.

\section{Discussion}

The inclusion of the finite $p_T$ contributions modifies the kinematics
of the problem somewhat.  Previous analyses have assumed that a $2 \to 1$
subprocess dominates the $J/\psi$ production.  If this is the case, it is
apparent that a fixed $cm$ energy ($\sqrt{s}$) and a fixed invariant mass
($M^{2}$), implies a fixed $\tau = M^{2}/s$.  But since
$\tau = x_{1} x_{2}$, the relations between the kinematical variables is
given by
$x_{1}=(\sqrt{4\tau + x_{F}^{2}} \ + x_{F})/2$,
$x_{2}=(\sqrt{4\tau + x_{F}^{2}} \ - x_{F})/2$.  Simply putting
$\tau = M_{J/\psi}^2/s$ and measuring $x_F$, one can extract the parton
momentum fractions.  Now, however, with the possible addition of more
final state partons, the invariant mass of the produced state is no
longer the $J/\psi$ mass, and so the kinematical relations must be
modified somewhat.  In Fig.~6, we show the invariant mass distribution
for $J/\psi$ and $\Upsilon$.  The average invariant mass in $J/\psi$
production is over 1~GeV above the $J/\psi$ mass (when we include only
the $g + g \to J/\psi + g$ subprocess), but the difference is relatively
smaller in the $\Upsilon$ case, so use of the correct kinematical
expressions for $x_1$ and $x_2$ is in order.  The inclusion of radiative
$\chi_J$ decays will not significantly alter the preceding argument.  The
correct expressions for $x_1$ and $x_2$ depend on the mass of the
produced charmonium state, it's $p_T$ and energy measured in the lab
frame, and $x_F$.  The expression can be simply derived starting from
equation (3.2) in Ref.~\cite{berger}
\begin{equation}
\hat{s} = \frac{M^2}{z} + \frac{p_T^2}{z(1-z)}
\end{equation}
where $z$ = $E_{onia}/E_{g_1}$ with the energies measured in the lab
frame and $g_1$ is the gluon from the beam.  Replace $z$ with
$z_{obs}/x_1$ ($z_{obs} = E_{onia}/E_{beam}$) and $\hat{s}$ with
$x_1 x_2 s$ and solve for $x_1$ and $x_2$ in terms of experimentally
measurable quantities.
\begin{eqnarray}
x_{1,2} & = & \frac{1}{2} \left\{ \left[ (x_F + z_{obs})
+ \frac{M_T^2}{s z_{obs}} \right] \right. \nonumber \\
& \pm & \left. \sqrt{ (x_F + z_{obs})^2
+ \left[ \frac{M_T^2}{s z_{obs}} \right]^2
- \frac{2 [x_F M_T^2 - z_{obs} (p_T^2 - M^2)]}{s z_{obs}}} \right\}
\end{eqnarray}
where $M_T = \sqrt{p_T^2 + M^2}$ is the transverse mass of the charmonium
state produced.  It is important to note that the values for $x_1$ and
$x_2$ assuming fixed $\tau$ are not even correct at $p_T = 0$, since it
is possible for the final state parton to be collinear with the charmonia
state without being soft.

The small-$x$ region represents one of the last frontiers of perturbative
QCD~\cite{collins}.  For this reason as well as for the very important
study of gluon shadowing more data, both at lower $x$ and with higher
statistics, are needed.

\vspace*{0.4cm}
{\Large{\bf \noindent  Acknowledgements}}
\vspace*{0.4cm}

One of the authors (MBGD) acknowledges the hospitality of the Institute
for Elementary Particle Physics Research at the University of Wisconsin
and the support by CAPES (Brazil).  This work was supported in part by
the U.~S. Department of Energy under Contract No.~DE-AC02-76ER00881, in
part by the Texas National Research Laboratory Commission under Grant
Nos.~RGFY9173 and RGFY9273, and in part by the University of Wisconsin
Research Committee with funds granted by the Wisconsin Alumni Research
Foundation.

\newpage

\begin{table}[h]
\begin{tabular}{c|c|c|c|c|c|c}  \hline \hline
Experiment & Beam   & Beam               & Target    & $x_2$
& $x_F$       & Reference                      \\
           & energy & type               &           & range
& range       &                                \\
           & (GeV)  &                    &           &
&             &                                \\ \hline \hline
NA3        & 43     & $\pi^-$            & Be, Cu, W & 0.1 - 0.3
& 0 - 0.95    & \cite{antipov}                 \\ \hline
NA3        & 39.5   & $\pi^-$            & H, W      & 0.11 - 0.34
& 0 - 0.85    & \cite{corden}                  \\ \hline
           & 150    & $\pi^-$            & H, Pt     & 0.03 - 0.17
& 0.01 - 0.9  &                                \\
NA3        & 200    & $\pi^-$, $p$       & H, Pt     & 0.0225 - 0.15
& 0.01 - 0.8  & \cite{badier}                  \\
           & 280    & $\pi^-$            & H, Pt     & 0.016 - 0.127
& 0.01 - 0.9  &                                \\ \hline
E537       & 125    & $\bar{p}$, $\pi^-$ & W, Cu, Be & 0.04 - 0.20
& 0.02 - 0.75 & \cite{katsanevas}              \\ \hline
E672       & 530    & $\pi^-$            & C, Al,    & 0.013 - 0.124
& 0.1 - 0.8   & \cite{kartik}                  \\
           &        &                    & Cu, Pb    &
&             &                                \\ \hline
E772       & 800    & $p$                & D, C, Ca, & 0.01 - 0.4
& 0.15 - 0.65 & \cite{alde1}                   \\
           &        &                    & Fe, W     &
&             & ($J/\psi$, $\psi'$)            \\ \hline
E772       & 800    & $p$                & D, C, Ca, & 0.1 - 0.35
& 0 - 0.7     & \cite{alde3}                   \\
           &        &                    & Fe, W     &
&             & ($\Upsilon$)                   \\ \hline
E705       & 300    & $p$, $\bar{p}$,    & Li        & 0.015 - 0.122
& 0 - 0.45    & \cite{antoniazzi}              \\
           &        & $\pi^+$, $\pi^-$   &           &
&             &                                \\ \hline \hline
NMC        & 280    & $\mu$              & H, D      & 0.02 - 0.3
& NA          & \cite{allasia}                 \\ \hline
NMC        & 200    & $\mu$              & C, Sn     & 0.02 - 0.2
& NA          & \cite{amaudruz1}               \\
NA37       & 280    &                    &           &
&             &                                \\ \hline
\end{tabular}
\caption[Summary of experimental results on $J/\psi$ production on
nuclear targets.  The values of $x_2$ were derived using
$\tau = M_{J/\psi}^2/s$.]{Summary of experimental results on $J/\psi$
production on nuclear targets.  The values of $x_2$ were derived using
$\tau = M_{J/\psi}^2/s$.}
\end{table}

\newpage

\noindent
{\bf \Large{Figure Captions}} \\

{\bf Figure 1} - Production of $J/\psi$ for a $\mu$ beam of energy
280~GeV on tin ($Sn$) and carbon ($C$) targets (experimental data from
Ref.~\cite{amaudruz1}).  The curves correspond to shadowing ($x_c = 0.03$
and $K_g = 0.50$), EMC effect and strong screening (dotted), shadowing
($x_c = 0.10$ and $K_g = 0.20$ for comparison), EMC effect and strong
screening (dashed), shadowing ($x_c = 0.03$ and $K_g = 0.50$) and strong
screening (dotdashed) and strong screening alone (solid).

{\bf Figure 2} - Production of $J/\psi$ (a) and $\psi'$ (b) for a proton
beam of energy 800~GeV on tungsten ($W$) and hydrogen ($H_2$) targets
(experimental data from Ref.~\cite{alde1}).  In Figure~2a, the best fit
requires the shadowing parameters $x_c = 0.185$ and $K_g = 0.04$, while
in Figure~2b, the best fit requires $x_c = 0.25$ and $K_g = 0.10$ (solid
line).  Also shown Figure~2b is the fit for $x_c = 0.20$ and
$K_g = 0.135$ (dashed line).

{\bf Figure 3} - Production of $J/\psi$ for a proton beam of energy
200~GeV on platinum ($Pt$) and $H_2$ targets (experimental data from
Ref.~\cite{badier}).  The curve shown includes only final state effects
and the `classical EMC' effect.

{\bf Figure 4} - Production of $J/\psi$ for a $\pi^-$ beam of energy
280~GeV on $Pt$ and $H_2$ targets (experimental data from
Ref.~\cite{badier}).  The best fit is achieved by including only final
state effects and the `classical EMC' effect.

{\bf Figure 5} - Production of $J/\psi$ for a $\pi^-$ beam of energy
200~GeV on $Pt$ and $H_2$ targets (experimental data from
Ref.~\cite{badier}).  The best fit is achieved by including only final
state effects and the `classical EMC' effect.

{\bf Figure 6} - The cross section {\it vs.} invariant mass
($\sqrt{\hat{s}}$) for the subprocess $g + g \to J/\psi + g$, assuming a
proton beam of energy 800~GeV on a $H_2$ target.

\end{document}